\documentclass[12pt,journal,draftclsnofoot,onecolumn]{IEEEtran}
\usepackage{fancyhdr}
\usepackage{amsmath,amssymb}
\usepackage{amsfonts}
\usepackage[dvips]{graphicx}
\usepackage{dsfont}
\usepackage{balance}
\usepackage{algpseudocode}
\usepackage{algorithm}
\usepackage{tikz}
\usepackage[hidelinks]{hyperref}    
\usepackage{color}
\usepackage{pgfplots}           
\pgfplotsset{compat=1.16}
\usepackage{graphicx}
\usepackage{subfigure}
\usepackage{cite}
\usepackage{mathrsfs}
\usepackage{stfloats}
\usepackage{tikz}
\usepackage{mathdots}
\usepackage{yhmath}
\usepackage{cancel}
\usepackage{color}
\usepackage{siunitx}
\usepackage{array}
\usepackage{xcolor}
\usepackage{multirow}
\usepackage{textcomp}
\usepackage{gensymb}
\usepackage{tabularx}
\usepackage{booktabs}
\usetikzlibrary{fadings}
\usetikzlibrary{patterns}
\usetikzlibrary{shadows.blur}
\usepackage{enumitem}

\begin{document}
\pagenumbering{arabic}
\setcounter{page}{1}
\title{Wireless-Fed Pinching-Antenna Systems (Wi-PASS) for NextG Wireless Networks
\author{Kasun R. Wijewardhana, Animesh Yadav, Ming Zeng, Mohamed Elsayed, Octavia A. Dobre, and Zhiguo Ding

\thanks{K. R. Wijewardhana, M. Elsayed, and O. A. Dobre are with the Faculty of Engineering and Applied Science, Memorial University, St. John’s, NL, Canada. A. Yadav is with the School of EECS, Ohio University, Athens, OH, USA. M. Zeng is with the Department of Electrical and Computer Engineering, Laval University, Quebec, QC, Canada. Z. Ding is with both the University of Manchester, UK and Khalifa University, UAE. This work was supported by the Canada Research Chairs Program CRC-2022-00187
and Natural Sciences and Engineering Research Council of Canada (NSERC), Discovery Program, Grant RGPIN-2019-04123. 
}}
}
\maketitle

\begin{abstract}
Waveguide-based pinching-antenna systems (PASS) have recently emerged as a promising solution to mitigate severe propagation losses in millimeter-wave and terahertz bands by intelligently and flexibly establishing line-of-sight links. However, their reliance on wire-based feeding confines deployment to areas near the base station (BS), limiting installation flexibility and making them cost-ineffective for serving distant users or regions. To overcome this challenge, this article proposes wireless-fed pinching-antenna systems (Wi-PASS), which employ wireless feeding to energize waveguides. Wi-PASS offer a practical and cost-efficient means to extend coverage beyond the BS vicinity. Several indoor and outdoor use cases demonstrate Wi-PASS advantages over PASS. Numerical results further show that Wi-PASS deliver higher data rates than conventional fixed-antenna systems, confirming the superior feasibility and performance of Wi-PASS. Key future research directions are also discussed to advance Wi-PASS deployment.
\end{abstract}

\section{Introduction} 

Next-generation (NextG) wireless systems are evolving to deliver immersive data rates, massive connectivity, high reliability, and ultra-low latency required by emerging applications, such as the metaverse, smart transportation, and digital twins. Meeting these demands necessitates the use of higher-frequency spectrum, including millimeter-wave (mmWave) and terahertz (THz) bands. However, signals at such frequencies suffer from severe propagation losses due to path loss, shadowing, and various forms of attenuation \cite{pathLoss}. 
A promising strategy is to exploit these bands when line-of-sight (LoS) conditions are available. Reconfigurable intelligent surfaces (RISs) \cite{RIS} and flexible antennas, such as fluid antennas \cite{fluidAntennas} and movable antennas \cite{movableAntennas}, have emerged as enabling technologies to establish favorable channel conditions. RIS can create virtual LoS paths by reflecting signals around obstacles, thereby mitigating LoS blockage. On the other hand, fluid and movable antennas locally reconfigure transceivers to adapt to spatial channel variations caused by small-scale multipath fading, enhancing the effective channel gain. 
Despite their benefits, RIS and flexible antennas face limitations. RIS suffers from double attenuation along the source–RIS–destination path and requires computationally intensive optimization that scales with the array size. Flexible antennas are effective against small-scale fading but have a limited impact on large-scale fading in the presence of LoS blockage. 

Recently, waveguide-based pinching-antenna systems (PASS) have been proposed as a viable alternative \cite{docomo2022,yang2025}. PASS deliver signals first through dielectric waveguides and then radiate energy using pinching-antennas (PAs)---flexible dielectric strips attached to the waveguide that function as leaky-wave antennas \cite{liu2025pinching}. Connected to the base station (BS) via a wired link, PASS can establish strong LoS channels by positioning PAs close to users \cite{yang2025}. They also allow efficient reconfiguration by flexibly attaching or removing PAs, thereby optimizing spatial power distribution \cite{zhiguo2024}.

However, PASS deployment is constrained by their wire-based feed, which restricts coverage to areas near the BS. Extending PASS to distant users requires long dielectric waveguides or cables, as they are expensive, fragile, and difficult to install in complex environments with buildings, terrain variations, or other obstacles. Waveguides may often be curved around barriers, introducing further power losses at bends, and their infrastructure demands costly conduits similar to fiber networks. Moreover, the limited coverage of individual PAs necessitates multiple BS-to-waveguide connections, further increasing cost and deployment complexity.

To extend PASS benefits without prohibitive infrastructure investment, we propose wireless-fed pinching-antenna systems (Wi-PASS), where the waveguide is fed by a relay antenna instead of a wired link. The relay antenna, positioned to maintain a LoS connection with the BS, wirelessly powers the PASS, enabling flexible deployment in areas far from the BS. This design preserves the advantages of PASS while eliminating costly and fragile end-to-end waveguide infrastructure.

The objective of this article is to introduce the Wi-PASS framework and demonstrate their potential for extending the PASS deployment in diverse scenarios. The main contributions are as follows:
\begin{itemize}
\item Present fundamentals of PASS with a brief review of recent developments in multiple access and emerging applications.
\item Propose the Wi-PASS architecture, highlighting its advantages and representative use cases in indoor and outdoor environments.
\item Validate the performance of Wi-PASS through numerical results and identify key future research directions to enable their integration into NextG wireless systems.
\end{itemize}

\section{Pinching-Antenna Systems (PASS)} 
\subsection{Preliminaries of PASS}
PASS provide a novel approach to reconfigure fading wireless channels by dynamically controlling antenna placement and configuration, as demonstrated by NTT DOCOMO \cite{docomo2022}. PASS employ a dielectric waveguide, typically made of materials such as polytetrafluoroethylene, as the transmission medium. Small dielectric elements, called PAs, are attached along the waveguide and serve as passive transmit/receive antennas. Each PA can be independently activated, deactivated, or repositioned along the waveguide via a mechanical system, allowing adaptive responses to changing channel conditions. With waveguides extending tens of meters, this enables flexible communication zones, LoS link formation by obstacle avoidance, and reduced path loss through proximity to users \cite{LoSblockage}. PASS also support scalable and flexible multiple-input multiple-output implementation by simply adding or removing PAs.

The PASS topology may include single or multiple waveguides, each connected to the BS via a dedicated radio frequency chain. The single waveguide topology is depicted in Fig.~\ref{system model} (on the left-hand side). PAs along the same waveguide radiate the same waveform with different phase shifts induced by propagation delays, implementing a passive beamforming strategy termed \emph{pinching beamforming}. Consequently, antenna placement is critical for performance, and it can be continuously adjusted using mechanical systems or pre-installed at fixed locations with selective activation to meet communication requirements.

\subsection{State-of-the-Art of PASS}
Optimizing PA placement is key to achieving LoS connectivity and mitigating small-scale fading. Research has focused on efficient resource allocation under hardware constraints and quality-of-service (QoS) requirements \cite{ming2025_resourceAllocation}. Single-waveguide PASS have explored multiple access techniques, including power-domain non-orthogonal multiple access and time division multiple access, to serve multiple users \cite{Zeng_WCL25}, whereas multi-waveguide deployments enable joint pinching and digital beamforming to mitigate inter-user interference \cite{bereyhi2025mimo}.

Beyond communication, PASS offer advantages for integrated sensing and communication (ISAC). Near-field characteristics can be exploited by using distributed PAs to enhance sensing resolution and accuracy \cite{zhang2025isac}, while energy-efficient implementation reduces cost. PASS also yield physical-layer security benefits: waveguide-based transmission and near-user antenna placement lower information leakage risk, and pinching beamforming can simultaneously strengthen channels to legitimate users and suppress power at eavesdroppers, improving secrecy capacity \cite{sun2025plc}.

\section{Proposed Wireless-Fed Pinching-Antenna Systems (Wi-PASS)}

\begin{figure*}
    \centering
    \includegraphics[width=1\textwidth]{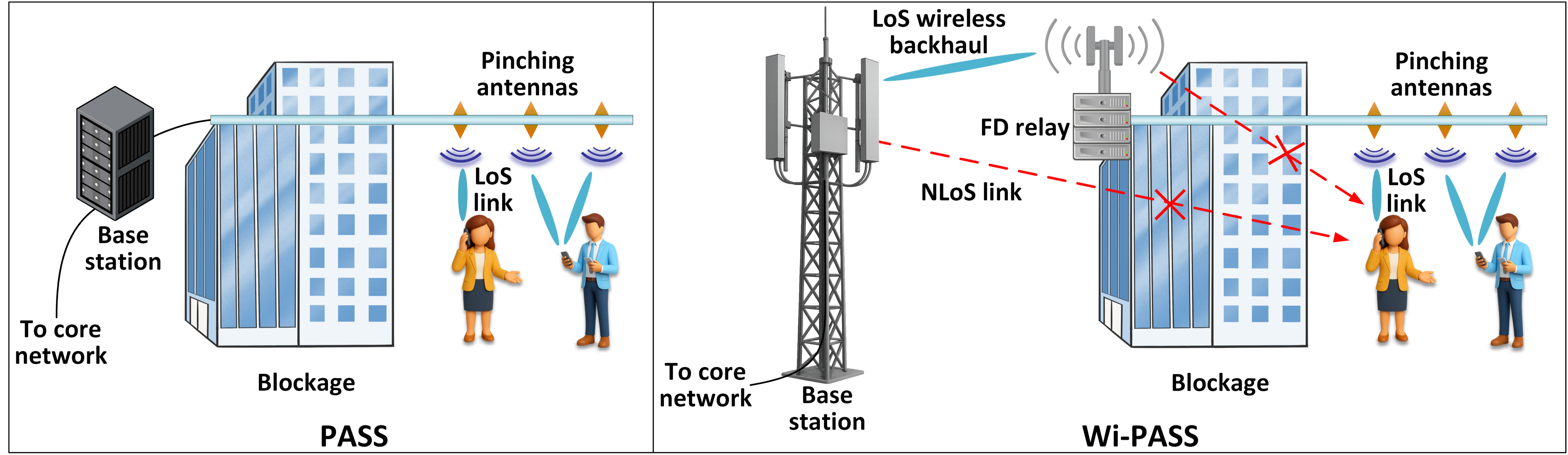}
    \caption{PASS versus Wi-PASS system architectures.}
    \label{system model}
\end{figure*}

Current PASS are confronted with two significant challenges: first, their deployment is typically confined to waveguides within a few tens of meters from the BS; second, given that the BS is linked to the core network, extending the core network to each PASS deployment incurs considerable costs and practical difficulties. Consequently, these limitations substantially hinder the system's overall potential. To extend their applicability and unlock their full capabilities, we propose a novel Wi-PASS architecture that integrates
relay functionality with PASS, enabling enhanced performance and practical deployment. As shown in Fig.~\ref{system model} (on the right-hand side), in Wi-PASS, a conventional antenna receives the BS signal, which is then amplified and forwarded into the waveguide in a full-duplex (FD) relay architecture, with the PASS acting as the transmitting antenna. The key idea is to replace long-wired feeds or extended waveguides with a wireless link, offering a practical and cost-effective solution for serving users far from the BS.

\begin{figure*}
    \centering
    \includegraphics[width=1\textwidth]{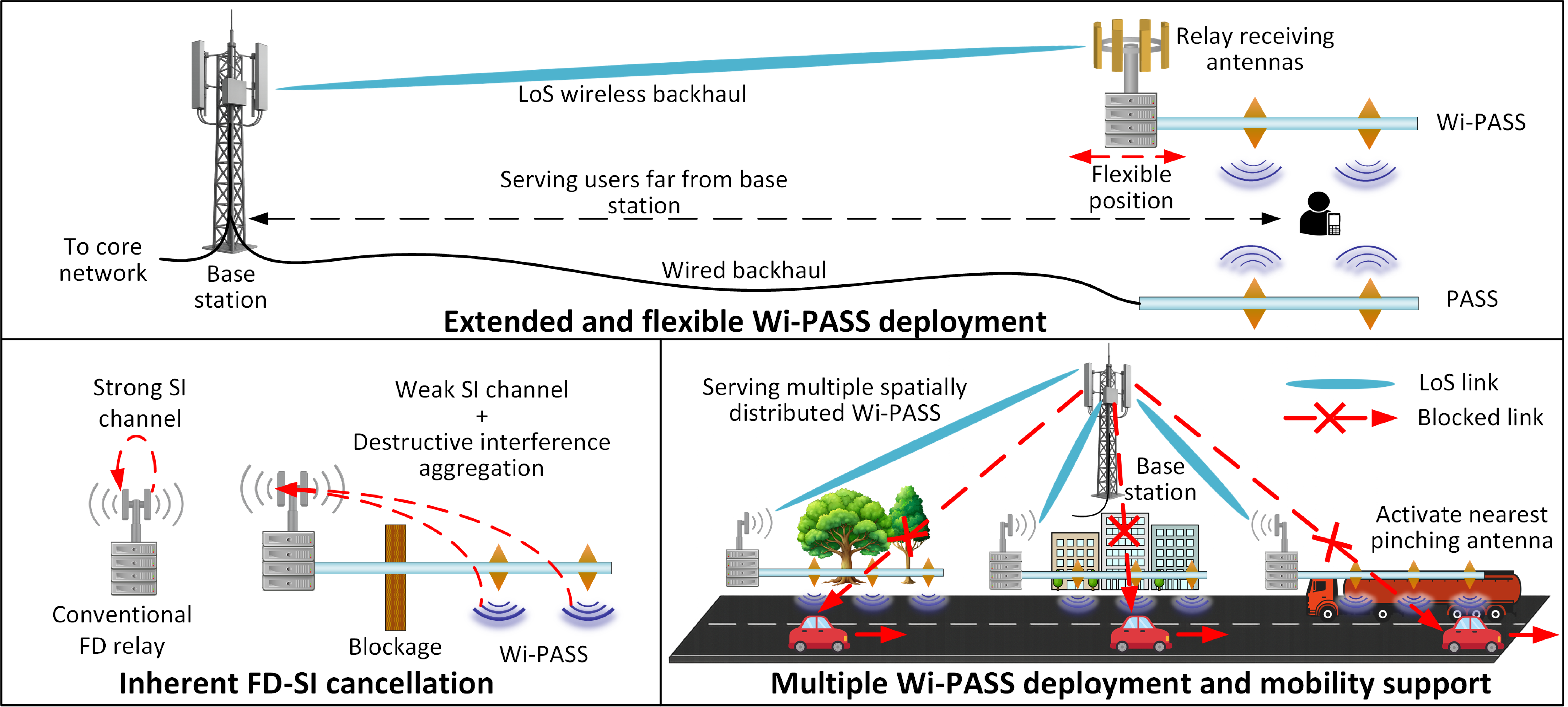}
    \caption{Key advantages of Wi-PASS deployment.}
    \label{PARS-advantages}
\end{figure*}

In addition to PASS ability to create LoS links and adapt antenna placement, Wi-PASS introduce several distinct advantages, as illustrated in Fig. \ref{PARS-advantages}:
\begin{itemize}
    \item \textbf{Extended range:} Wireless feeding allows PASS to be deployed far from the BS, eliminating wired backhaul and reducing cost and installation complexity.
    \item \textbf{Flexible deployment:} Relays can be flexibly positioned in challenging or remote environments, extending coverage and capacity where conventional infrastructure is impractical.
    \item \textbf{Multi-PASS support:} A BS can serve multiple distributed relays, enabling deployment of several PASS units from a single BS and improving infrastructure utilization.
    \item \textbf{Mobility support:} Multiple Wi-PASS nodes placed along a user’s path enable seamless handover to the nearest node, maintaining strong LoS links and reliable connectivity for high-mobility users.
    \item \textbf{Inherent FD-SI mitigation:} By physically separating the relay antenna from the PAs, Wi-PASS increase the path loss of the self-interference (SI) channel. Strategic PA placement can further induce destructive interference, reducing SI power and improving performance.
\end{itemize}

Like PASS, Wi-PASS can accommodate diverse waveguide and PA configurations. The relay receiving antenna can also be implemented as PASS, exploiting their advantages for signal reception. It is important to note that wireless backhaul typically has lower capacity than its wired counterpart; this limitation in Wi-PASS can be mitigated by utilizing the vast bandwidth available in mmWave and THz bands.

\section{Use Cases: Diverse Wi-PASS Deployment Scenarios}
\begin{figure*}
    \centering
    \includegraphics[width=1\textwidth]{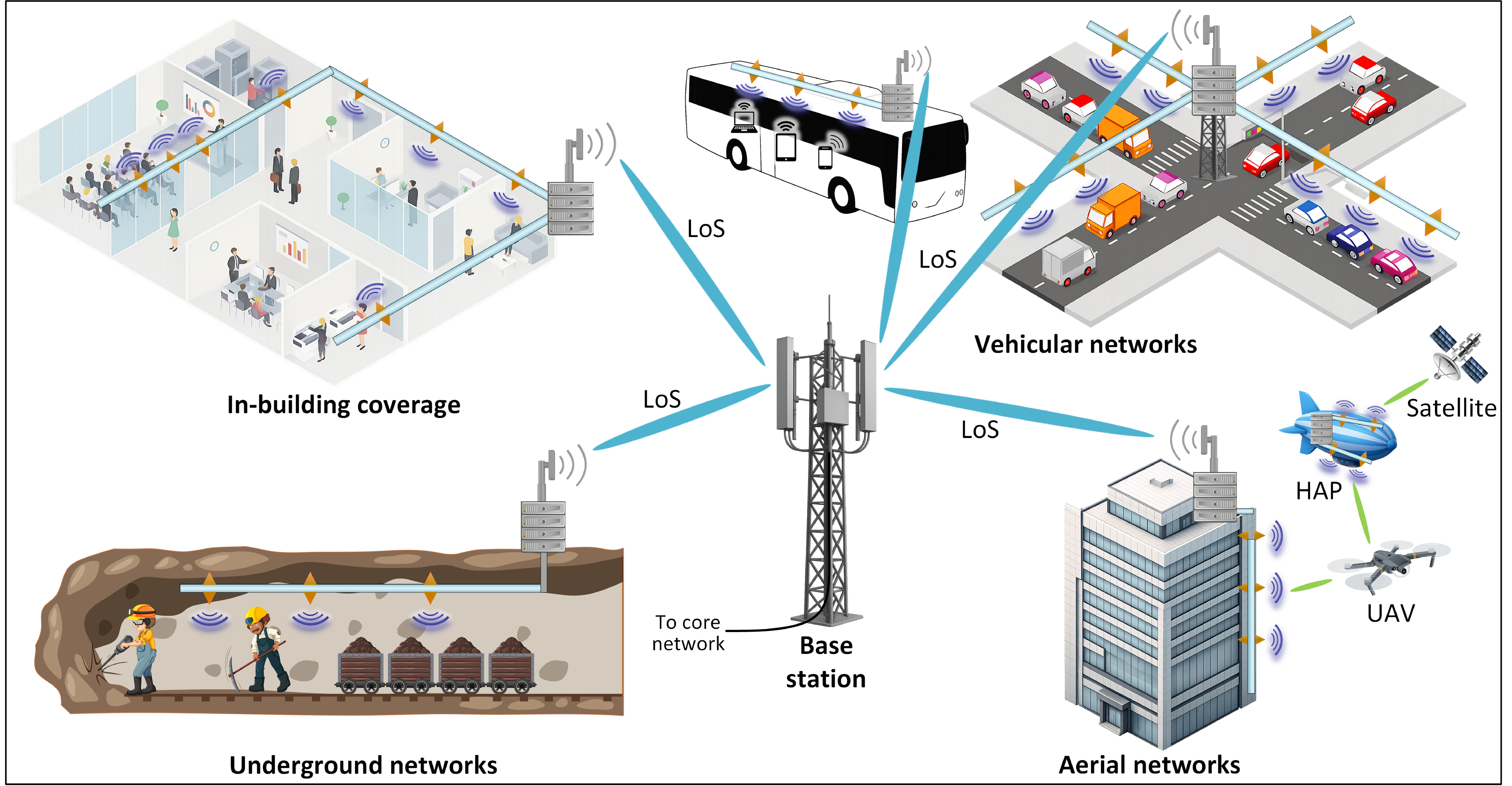}
    \caption{Use cases of Wi-PASS in diverse communication scenarios.}
    \label{use cases}
\end{figure*}

With flexible and adaptive deployment, Wi-PASS enable diverse applications in both indoor and outdoor environments, as shown in Fig. \ref{use cases}. This section explores deployment scenarios, where Wi-PASS can overcome the inherent limitations of conventional PASS and existing fixed-antenna systems.

\subsection{Wi-PASS for Indoor Networks}
\subsubsection{In-Building Coverage}
Indoor wireless communication is often impaired by penetration losses in walls and multipath propagation due to furniture and other obstacles. Conventional relays and repeaters provide limited coverage and data rates in such environments. Wi-PASS offer a more effective solution by placing the receiving antenna outside to capture stronger signals and then routing the dielectric waveguide indoors along ceilings. This allows the BS to serve multiple indoor spaces without dedicated wired links. PAs can then be strategically deployed to create localized LoS zones, reducing scattering, interference, and electromagnetic pollution. Moreover, as indoor layouts frequently change, PASS enable flexible repositioning of antennas without extensive hardware upgrades, offering a cost-effective alternative to Wi-Fi or repeater-based systems.

\subsubsection{Underground Networks}
Underground environments, such as tunnels, subways, and mines, pose severe challenges due to severe shadowing, narrow spaces, and harsh communication conditions. Conventional wireless systems suffer from power losses, interference, and limited reliability, while wired solutions are costly to deploy and maintain. Wi-PASS mitigate these issues by placing receiving antennas above ground and routing PASS underground through drilled shafts. The dielectric waveguide can be laid along the tunnel or mine, with relay receivers deployed periodically above ground to extend coverage. In dynamic settings, such as active mines, PAs can be repositioned to adapt to changing layouts, unlike conventional systems that require new infrastructure. Furthermore, dielectric materials in PASS are resilient to humidity, dust, and temperature fluctuations, while critical electronics remain above ground, ensuring robust operation in harsh underground conditions.

\subsection{Wi-PASS for Outdoor Networks}

\subsubsection{Vehicular Networks}
Vehicular networks require ultra-reliable, low-latency communication to support intelligent transportation and autonomous driving networks. Obstacles along the road infrastructure, such as high-rising buildings, trees, and large vehicles on the road can cause frequent LoS blockage between BS and the vehicle. Wi-PASS help to address LoS blockage, intermittent connectivity, and frequent handovers caused due to high mobility by enabling LoS links and adaptively activating the PA closest to the moving vehicle. However, deploying a single long PASS along roadways is impractical due to excessive propagation losses and high costs for installation and maintenance. A more feasible approach is to deploy multiple Wi-PASS nodes with moderate waveguide lengths distributed along roadsides. These nodes can be fed via wireless links, eliminating the need for direct wired connections to the BS and simplifying deployment in remote areas. Moreover, Wi-PASS can enhance in-vehicle connectivity by deploying the relay receiving antenna on the vehicle roof while positioning the PASS on the interior ceiling.

\subsubsection{Aerial Networks}
Aerial platforms, such as uncrewed aerial vehicles (UAVs), high-altitude platform stations (HAPS), and satellites, critically rely on LoS links to overcome high path loss associated with long distances and mobility. PASS mounted on rooftops or building facades can support such links, but Wi-PASS are preferred since wired backhaul on tall structures is costly and complex. In urban settings, UAVs typically fly at higher altitudes to maintain LoS above tall buildings, whereas Wi-PASS on building facades enable strong LoS connectivity at lower altitudes, reducing their energy consumption and improving link reliability. In disaster scenarios, elevated Wi-PASS nodes can remain functional even if the terrestrial infrastructure is damaged, ensuring reliable UAV support. Furthermore, the lightweight and flexible design of Wi-PASS---when both transmitter and receiver relays employ PASS---makes them particularly suitable for coordinated HAPS deployments where efficiency and adaptability are critical.

\section{Performance Studies}
This section evaluates the performance of the proposed Wi-PASS against end-to-end PASS and conventional fixed-antenna relay systems. For illustration purposes, a downlink single-user scenario is considered, with achievable data rate as the performance metric. Five transmission schemes are compared:
\begin{enumerate}
    \item \textbf{Proposed Wi-PASS}: The BS employs multiple antennas, while the relay uses a single-antenna receiver and operates with constant transmit power under FD-SI. 
    \item \textbf{PASS}: A single waveguide with one active PA, which is placed to minimize user distance \cite{zhiguo2024}.
    \item \textbf{FD relay (ideal)}: A single receiver antenna and multiple transmit antennas without FD-SI, assuming the relay-user LoS link is blocked.
    \item \textbf{FD relay (practical)}: A single receiver antenna and multiple transmit antennas with FD-SI, assuming the relay-user LoS link is blocked.
    \item \textbf{Conventional channel}: A direct BS-user transmission over a Rayleigh-faded channel.
\end{enumerate}

All schemes assume that the user locations are uniformly distributed within a square area, and results are obtained via Monte Carlo simulations. Simulation parameters are listed in Table \ref{tab:simPara}.

\begin{figure}
    \centering
    \includegraphics[width=0.5\textwidth]{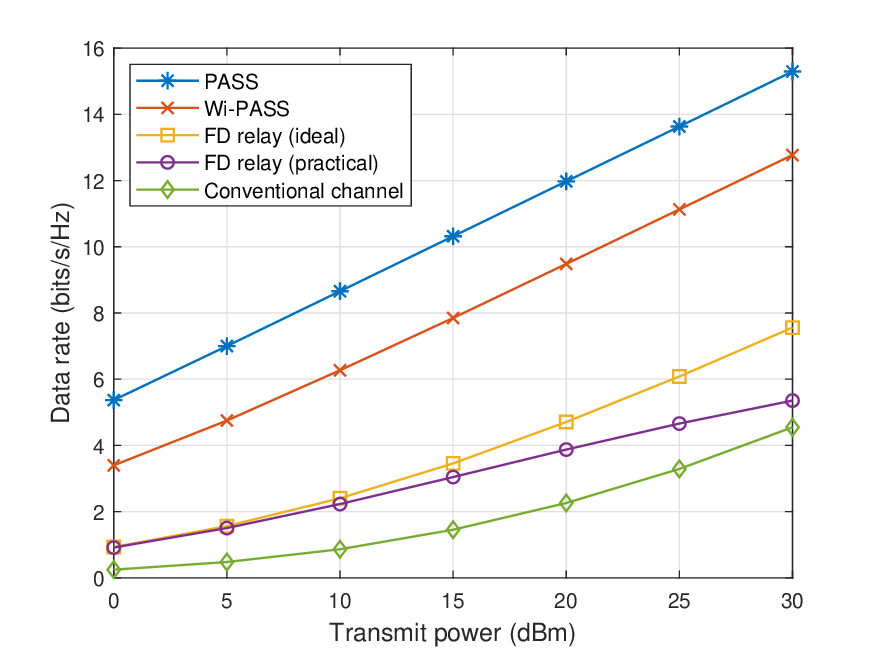}
    \caption{Data rate versus the transmit power for a BS-relay distance of 50 meters.}
    \label{fig1}
\end{figure}

Fig. \ref{fig1} shows the achievable data rate versus the BS transmit power. Both Wi-PASS and PASS consistently outperform fixed-antenna relays by exploiting near-user LoS links to reduce path loss. PASS achieve the highest rate due to minimal in-waveguide loss, but its reliance on wired feeding limits deployment flexibility and increases cost. Wi-PASS, while slightly inferior to PASS, outperform fixed-antenna relays and provide more practical deployment. Although FD operation introduces SI in Wi-PASS, antenna separation inherently suppresses SI, reducing the need for complex cancellation techniques. In contrast, conventional FD relays suffer severely from SI due to small antenna separation, requiring advanced suppression methods. The conventional channel performs the worst due to high path loss.

\begin{figure}
    \centering
    \includegraphics[width=0.5\textwidth]{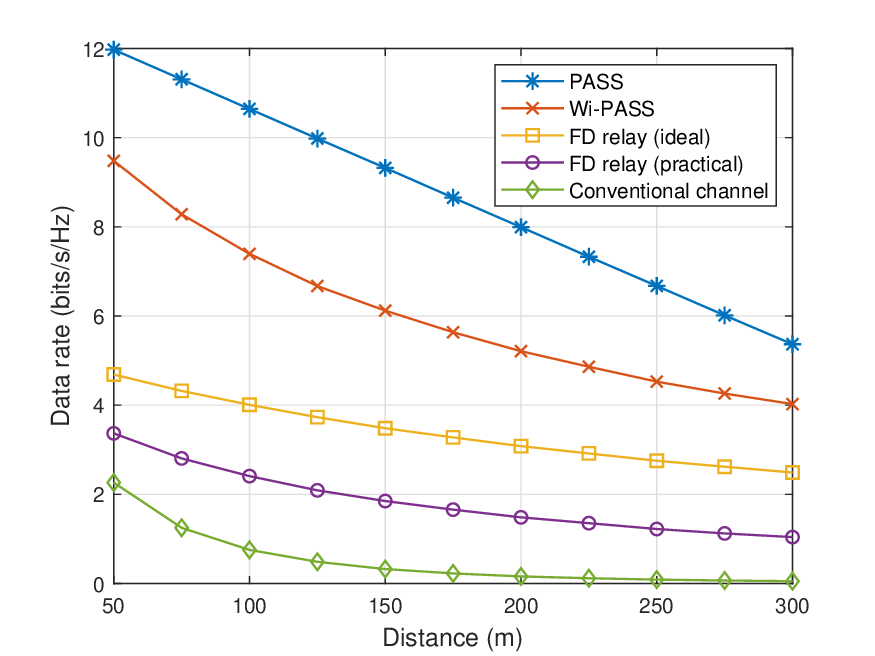}
    \caption{Data rate versus the BS-relay distance for a transmit power of 20 dBm at both the BS and Wi-PASS relay.}
    \label{fig2}
    \vspace{-4mm}
\end{figure}

Fig. \ref{fig2} examines the effect of BS–relay distance. The user is kept at a fixed distance from the BS across schemes. Both Wi-PASS and PASS outperform fixed-antenna relays over the entire range. PASS rates decline with distance due to accumulated in-waveguide loss, while in Wi-PASS, the reduced relay input power is partly compensated by amplification and the strong relay–user LoS link. Conventional relays remain limited by non-LoS conditions, and the conventional channel degrades rapidly at longer distances because of severe high-frequency path loss.

\begin{table}
    \centering
\caption{Simulation set up parameters.}
\label{tab:simPara}
    \begin{tabular}{|l|l|}\hline
        \textbf{Parameter} & \textbf{Value}\\\hline
        PASS height & 3 m\\\hline
        Coverage area dimension & 10 m $\times$ 10 m\\\hline
        BS-relay distance & 50 m\\\hline
        Carrier frequency & 28 GHz\\\hline
        Noise power & -90 dBm\\\hline
        Waveguide propagation loss & 0.08 dB/m\\\hline
        No. of transmit antennas & BS = 12, relay = 12\\\hline
        Path loss exponent & BS-relay/relay-user = 2.55, \\&BS-user = 4\\\hline
        Shadowing variance & 11 dB\\\hline
        Rician $K$ factor & 10 dB\\\hline
        Relay transmit power & [0:30] dBm\\\hline 
        BS transmit power & [0:30] dBm\\\hline
        FD-SI cancellation level & -85 dB\\ \hline
    \end{tabular}
\end{table}

\section{Future Research Directions of Wi-PASS}
This section discusses key research directions for future Wi-PASS deployment.

\subsection{Resource Allocation}
The use of wireless backhaul in Wi-PASS extends PASS coverage by enabling relay deployment at diverse locations served by the same BS. However, each Wi-PASS deployment experiences distinct fading channels and FD-SI, leading to variations in received power and interference. Meeting QoS requirements thus necessitates efficient power allocation strategies at the BS and/or relay. PA placement is a key challenge due to transmit power limits and waveguide propagation losses, requiring energy-efficient activation and positioning mechanisms. For multiuser scenarios, waveguide placement optimization enables LoS short-range links and introduces additional degrees of freedom, demanding computationally efficient resource allocation strategies. Developing such methods is essential to fully exploit Wi-PASS in NextG wireless systems.

\subsection{Hardware Design}
Wi-PASS can be integrated into existing network architectures with minimal modification, but still requires additional components such as FD relays. System design must consider relay processing capacity, power budget, complexity, and cost. Hardware impairments and waveguide propagation losses constrain both waveguide length and the number of PAs. Flexible positioning also depends on precise mechanical adjustments, which increase control complexity, mechanical wear, latency, and precision issues. Multiport network theory \cite{liu2025tutorial} provides a framework for designing equivalent circuits to emulate PAs, addressing feasibility concerns. Tackling these hardware challenges is critical for ensuring reliable Wi-PASS performance.

\subsection{Channel Estimation}
Accurate channel state information (CSI) is essential for PA placement to enable constructive aggregation at the receiver. Existing CSI estimation methods assume fixed antennas and far-field models, making them unsuitable for PASS/Wi-PASS, where near-field effects and flexible positioning dominate. Conventional approaches also incur excessive pilot and hardware overhead. In Wi-PASS, CSI estimation is further complicated by FD-SI, requiring simultaneous estimation of relay-to-user and SI channels. We note that Wi-PASS enables reconfigurable antenna separation, which introduces challenges to SI channel estimation and affects interference cancellation. Since accurate CSI directly impacts power allocation, and Wi-PASS nodes have limited power and computational capacity, efficient low-overhead CSI estimation methods tailored to Wi-PASS are needed.

\subsection{Cooperative Deployment}
Wi-PASS nodes can cooperate to improve reliability and spectral efficiency through cooperative diversity and pinching beamforming. In dense urban environments with frequent blockages, multi-hop transmission via cooperative Wi-PASS nodes enhances end-to-end performance through successive FD relaying. Unlike traditional relays, Wi-PASS reduce hop distances using waveguide propagation, lowering power consumption and enabling reconfigurable routing via antenna repositioning. Furthermore, cooperation between the BS and Wi-PASS facilitates joint transmission strategies in cellular networks. However, cooperative deployment requires joint signal processing and resource allocation, making the design of efficient cooperative architectures more challenging.

\subsection{Emerging Technologies}
Due to their superior compatibility and feasibility, Wi-PASS can significantly enhance emerging wireless systems. In ISAC, Wi-PASS can facilitate sensing range extension and accuracy enhancement via near-user LoS links. Distributed deployment with flexible antennas presents a promising direction to achieve mobile target sensing with uniform coverage. In space–air–ground integrated networks (SAGIN), Wi-PASS are capable of addressing challenges from heterogeneous SAGIN platforms (UAVs, HAPS,  low earth orbit satellites) through adaptive deployment. Co-deployment with RIS and other reconfigurable antenna systems enables joint manipulation of the wireless channel, providing complementary benefits. Investigating Wi-PASS-aided architectures is thus critical to unlock performance gains across multiple subsystems.

\section{Conclusions}
This article introduced Wi-PASS as a practical approach for deploying PASS beyond the immediate vicinity of the BS. We reviewed PASS fundamentals, highlighted Wi-PASS advantages, and presented promising use cases in indoor and outdoor scenarios. Numerical evaluations confirmed Wi-PASS performance gains over fixed-antenna schemes in terms of downlink data rate. Finally, several key research directions were outlined, including resource allocation, hardware design, channel estimation, cooperative deployment, and integration with emerging technologies, to advance Wi-PASS adoption in NextG wireless networks.

\bibliographystyle{IEEEtran} 
\bibliography{bibliography.bib}
\vspace{-20 mm}
\begin{IEEEbiographynophoto}{\textsc{Kasun R. Wijewardhana}}
     [S] (gwijewardhan@mun.ca) is currently pursuing his master’s degree in computer engineering at Memorial University, Canada.  He received his B.Sc. degree in electrical and electronic engineering from the University of Peradeniya, Sri Lanka, in 2017. He was a radio network planning engineer at Dialog Axiata PLC, Sri Lanka, until 2024.
\end{IEEEbiographynophoto}
\vspace{-20 mm}
\begin{IEEEbiographynophoto}{\textsc{Animesh Yadav}}
     [SM] (yadava@ohio.edu) is an Assistant Professor in the School of Electrical Engineering and Computer Science at Ohio University, USA. His research interests include 6G and beyond networks, communications and networking for advanced air mobility (AAM) systems. He is a Senior Member of IEEE and serves as a Senior Editor for IEEE Communications Letters, an Associate Editor for Frontiers in Communications and Networks. Dr. Yadav received his master’s degree from Indian Institute of Technology (IIT) Roorkee, India, and Ph.D., degree from the University of Oulu, Finland.
\end{IEEEbiographynophoto}
\vspace{-20 mm}
\begin{IEEEbiographynophoto}{\textsc{Ming Zeng}}
     [M] (ming.zeng@gel.ulaval.ca) received the B.E. and master’s degrees from the Beijing University of Posts and Telecommunications, Beijing, China, in 2013 and 2016, respectively, and the Ph.D. degree in telecommunications engineering from the Memorial University of Newfoundland, St John’s, NL, Canada, in 2020. He is currently an Associate Professor and the Canada Research Chair with the Department of Electrical and Computer Engineering, Laval University, Quebec, QC, Canada. He has published more than 100 articles and conferences in first-tier IEEE journals and proceedings, and his work has been cited over 5,400 times per Google Scholar. His research interests include resource allocation for beyond 5G systems and machine learning-empowered optical communications. He serves as an Associate Editor for IEEE Transactions on Communications, IEEE Open Journal of the Communications Society, and IEEE Wireless Communications Letters.
 \end{IEEEbiographynophoto}
\begin{IEEEbiographynophoto}{\textsc{Mohamed Elsayed}}
     [M] (memselim@mun.ca) received the B.Sc. degree in Electronics and Communications Engineering from Sohag University, Sohag, Egypt, in 2014, the M.Sc. degree in Electronics and Communications Engineering from Assiut University, Assiut, Egypt, in 2018, and the Ph.D. degree in Electrical Engineering from Memorial University, St. John’s, NL, Canada, in 2024. He is currently serving as a Postdoctoral Fellow with the Faculty of Engineering and Applied Science, Memorial University, St. John’s, NL, Canada. He is also on leave from his position as an Assistant Professor with the Faculty of Engineering, Sohag University, Egypt. His research interests include spatial modulation, full-duplex communications, pinching antennas, and the application of machine learning and quantum machine learning in wireless communications. He was the recipient of the Best Paper Award at the 35th National Radio Science Conference (NRSC) in 2018 and the IEEE Newfoundland and Labrador Graduate Scholarship in 2023.
\end{IEEEbiographynophoto}
\vspace{-120 mm}
\begin{IEEEbiographynophoto}{\textsc{Octavia A. Dobre}}
    [F] (odobre@mun.ca) is a Professor and Tier-1 Canada Research Chair at Memorial University, Canada. Her research spans wireless, optical, and underwater communications. She serves as VP Publications of the IEEE Communications Society and was the founding Editor-in-Chief of the IEEE Open Journal of the Communications Society, as well as former EiC of IEEE Communications Letters. Dr. Dobre is a Fellow of the Engineering Institute of Canada, the Canadian Academy of Engineering, and the Royal Society of Canada, as well as a Clarivate Highly Cited Researcher.
\end{IEEEbiographynophoto}
\vspace{-120 mm}
\begin{IEEEbiographynophoto}{\textsc{Zhiguo Ding}}
     [F] (zhiguo.ding@ku.ac.ae) is currently a Professor in Communications at the University of Manchester, a Distinguished Adjunct Professor at Khalifa University, and an Academic Visitor at Princeton University. His research interests are 6G networks, communications, and signal processing. His h-index is over 120, and his work receives 70,000+ Google citations. He is serving as an Area Editor for the IEEE TWC, TCOM and OJ-SP, and an Editor for IEEE TVT. He is a Web of Science Highly Cited Researcher in two disciplines (2019-2024), and a Fellow of the IEEE.
\end{IEEEbiographynophoto}

\end{document}